# ProcDSL + ProcEd - a Web-based Editing Solution for Domain Specific Process-Engineering

Christian Berger     Tim Gülke     Bernhard Rumpe

RWTH Aachen University
Software Engineering Group
Ahornstraße 55
52074 Aachen, Germany

www.se-rwth.de

## ABSTRACT

In a high-tech country products are becoming rapidly more complex. To manage the development process as well as to encounter unforeseen challenges, the understanding and thus the explicit modeling of organizational workflows is more important than ever. However, available tools to support this work, in most cases force a new notation upon the company or cannot be adapted to a given publication layout in a reasonable amount of time. Additionally, collaboration among colleagues as well as different business units is complicated and less supported. Since it is of vital importance for a company to be able to change its processes fast and adapt itself to new market situations, the need for tools supporting this evolution is equally crucial. In this paper we present a domain specific language (DSL) developed for modeling a company's workflows. Furthermore, the DSL is embedded in a web-based editor providing transparent access using modern web 2.0 technologies. Results of the DSL's as well as the editor's application to document, model, and improve selected workflows of a German automotive manufacturer are presented.

## 1. INTRODUCTION AND MOTIVATION

In today's world of business-processes, modeling becomes a vital factor in an organization's change and process management and even daily routines. This in particular holds for the development of complex machines, such as airplanes, cars or trains. Each of these domains has their own specific problems, e.g. induced through supplier integration, need for quality certification, development for individual customers or the mass market, etc.

It is therefore not surprising that there is no unique solution for the management of these processes. Therefore, it is only natural to find company-specific layouts of process-descriptions and publications in almost every firm. Unfortunately, tools meant to support organizations in planning and developing their processes, often force their own layout, notation and logic upon their users. Although this might be considered easier and even 'better' than what the company is used to, we found it is one main reason to see printed Microsoft PowerPoint slides and similar documents to arise all over office walls. Big organizations need a certain amount of time to agree on a specific appearance of their process-documents and even longer to publicize this throughout the company. And even worse, the meaning of icons or the position of images tends to change during a company's evolution. Programs like [6] or [2] are not able to be easily adapted to appear like what people know and have worked with already. This fact clearly shows the need for a modular tool which can be adapted with considerably less effort than any others available.

Furthermore, many currently available tools are single user applications with only limited possibility for company-wide collaboration. Using MontiCore [5], a framework for developing domain specific languages, we developed a web-based editor for modeling organizational workflows that uses a DSL's instances as input and output. This DSL was developed together with company-experts to ensure correctness and completeness. The editor's interface then was constructed separately, so there was a clean cut between the logic and its representation. This enables us to change either the logic behind a process-plan or the frontends' appearance without touching the other. Process modeling is then being performed by the end user through a web-browser to gain the amount of flexibility necessary in today's quickly changing world. AJAX technology enables us to construct an interface almost as powerful as a traditional application's one.

This paper is structured as follows. First, a brief overview of MontiCore is presented. Following, the design considerations and the implementation of a DSL to model organizational workflows are discussed. This DSL is embedded in a web-based process editor which is presented afterwards. Finally, the DSL as well as the editor's application is shown on an example from the automotive domain.

## 2. MONTICORE – A FRAMEWORK FOR DEVELOPING DOMAIN SPECIFIC LANGUAGES

MontiCore is a framework for developing textual domain specific languages from the Department of Software Engineering at RWTH Aachen University. It supports grammar-based language design as well as meta-modeling concepts by providing one language for abstract and concrete syntax definition. Using a given definition it generates automatically a lexer, a parser, and classes for an abstract syntax graph (ASG) describing the language's structure. At runtime, these classes represent a successfully parsed and processed instance of a given language [5, 7, 8, 9].

Generated artifacts and MontiCore itself are coded in Java. Due to its sophisticated language processing concepts and its support for Java which is also used by the technology we intended to use for realizing the web-based editor, we have chosen MontiCore for defining the language and for processing instances at runtime.

## 3. PROCDSL – A DSL FOR PROCESS DESCRIPTIONS

We propose a domain specific language to represent the company's organizational workflow processing. The reason we used a DSL to formalize the logic behind a process-plan was the complex structure of those plans, hidden behind a rather simple appearance. Basically, a milestone's appearance in a specific plan was determined by the organizational view, consisting of a layer and unit combination, the plan represented. One unit might be only participating to the milestone's result while another one is responsible for it. Both types of access are represented through different icons in their unit's process-plan.

Without the use of MontiCore and a DSL, we would only have been able to construct an application that suits the current needs and requirements as we understood them. In case of a sudden change of the appearance or logic of those process-plans, the application would have to be reconstructed in a time-consuming way. Through our separation it is now possible to change either the model or the graphical representation without touching the other. The DSL, which we modeled together with chosen experts, enabled us to already start the development of the editor-frontend while still being in the process of figuring out the logic's details behind the plans. This will also save resources later if for example different views will be needed for the same data. We predict that in near future, a plan's layout will change again or a new organizational layer might be implemented - in that case the model or the editor can be changed quickly without the need to rewrite a whole database scheme and an applications access to it.

The main advantage over pure visual process modeling tools is its formal specification. Additionally, a textually defined DSL can be simply embedded in different contexts and therefore easily reused. In the following, we present briefly the DSL we designed for modeling organizational workflows considering the following design criteria.

- *Intuitional Representation.* Instead of using XML for defining workflows we chose a much more simple representation to avoid XML's verbosity and redundancy in its data description. Thus, a better readability for the user can be achieved if DSL's instances are processed without a graphical editor.

- *Small Data Format.* Since the language is intended to be used in a web-based context, large entities of organizational workflow descriptions would cause lots of bandwidth consumption. Thus, a small data format to be exchanged with a server is desirable.

- *Reusability.* The language itself is primarily intended to be used with a graphical web-based editor to support process engineers. However, having a formally defined and application-independent process description, language's instances can be easily exchanged among the same application. Moreover, other tools can be used for checking semantic constraints on the one hand or to transform an instance into another data format on the other hand.

- *Versioning.* Regardless if an available solution like Subversion is used or a domain-specific (e.g. graphical) one is programmed it is obvious that textual formats are easier to put under version control as well to track and compare changes.

To ensure usefulness, domain experts from the company were heavily involved in the development of the DSL. Using a simple UML-representation of the DLS's structure, we were able to communicate in a productive way.

In Fig. 1, an excerpt of our grammar is shown. Technically, MontiCore accepts productions with EBNF-like right hand sides. Nonterminals (like `Milestone` or `String`) can be preceded by attribute names (like in `name:String`). Attribute names can also be attached to terminals like `"Scope"` or `"resp"` denoting, whether the keyword was detected.

Lines 1-5 contain the grammar's start symbol. The workflow description starts with a header containing some meta-information about the current instance followed by a list of milestones. Every milestone has a name, a description, and several other properties of which some are included in 1, lines 9-13. Line 11 positions the milestone relatively to a timeline. As already mentioned, the need to separate the logic behind a process-plan and it its actual graphical representation was crucial. Therefore, during development of the DSL, we made sure not to mix graphical information like icon positions, colors, and the like with logic-related things. As a result, an instance of the given DSL does not only represent a milestone-plan like the one shown in Fig.4 which was used as a blueprint for the DSL, it also enables developers to get different graphical representations out of it (e.g. simple lists of milestones, a specific view on inputs and outputs of a milestone or the involvement of a layer in process activities).

Besides an informal description, a milestone has a concrete result which can be any appropriate artifact depending on a specific workflow. Different scopes and layers can access

```
                    MontiCore-Grammar
 1  ProcessFile =
 2    "process"
 3    ProcessHeader
 4    :Milestone*
 5    Process
 6    :Scope*
 7    "end";
 8  ...
 9  Milestone = "milestone" Name
10  ...
11    "position" TimelinePosition:Number
12    "result" Result:Result*
13    "description" Description:String;
14  ...
15  Scope = "scope" Name
16    "description" Description:String
17    r:Responsibility*;
18
19  Responsibility = "responsibility"
20    (responsible:["resp"]
21    | contributing:["cont"]
22    | noticing:["noti"])
23    "asmilestone" asMilestone:STRING;
24  ...
25  associations {
26    Responsibility.milestone * -> 1 Milestone;
27  }
28
29  concept sreference {
30   ResponsibilityMilestone:
31     Responsibility.asMilestone = Milestone.name;
32  }
```

**Figure 1: Excerpt of our grammar to describe organizational workflows.**

a milestone in different ways, like being responsible or just contributing to the result. In this case, a scope is a specific organizational unit within a layer, like *manufacturing* within the layer *departments*. Combined, these selections define different views on the whole set of milestone-data. A scope's responsibilities are described in lines 19-23. Every scope is either *directly responsible* for fulfilling a sub-process associated with this milestone, *contributing* for a concrete milestone or only *noticing* the state of a sub-process.

Using the concept of automatically set associations provided by MontiCore in line 25-32, the responsibilities' milestones are navigably associated with an ASG node describing a milestone. The following lines starting at line 27 describe the way a milestone is mapped by its (unique) name to the corresponding responsibility-object's association.

For validating given values of concrete DSL's instances object which traverses the ASG, generated by MontiCore can be defined. For example, a time-validating visitor can be used to check the semantic constraints whether the start time of a given milestone is prior to its end time regarding to the underlying timeline specification, which can be either a regular calendar or a simple sequence of weeks.

Using the grammar outlined in this section, we designed and implemented a graphical web-based editor which is described in the following.

## 4. GRAPHICAL EDITOR FOR PROCDSL USING WEB 2.0 TECHNOLOGIES

We wanted the graphical editor to be as easily usable as possible combined with the flexibility a web-application gives us regarding deployment and maintenance. AJAX enables developers to design web-applications that make use of asynchronous callbacks rather than of synchronized ones. Therefore, the traditional request-response-paradigm is no longer the limiting factor in a web-application's interface. Using AJAX different parts of the website can be loaded dynamically providing a great range of possibilities to the developer to design the application. For more information about the AJAX technology see [12].

Since the overall layout was already fixed due to the fact that we were working with a company which had already specified its appearance for process descriptions, it was clear that the editor should not be a generic canvas, but an aid to work in that given layout. However, it should use the formalism provided with the DSL to keep users from inventing new icons and limit them to correct instances.

We selected the Google Web Toolkit[4] as our main framework which enabled us to write Java-code instead of JavaScript for the web-interface. Through this, a highly interactive web-based application combined with proper testing and a decent coding-style was possible with much less effort than a traditional one would have required. The implementation of Drag&Drop-capabilities as displayed in Fig.2 for canvas-objects as well as dialog-windows is another factor that makes the interface a lot more comfortable for users. Through asynchronous callbacks, drafts can be saved and restored automatically.

As one can see in Fig.3, the main window is divided in three different areas. The largest is used by the actual milestone-plan, while the other two keep a toolbox to drag objects out from onto the plan and an object-inspector. The latter allows users to look into a chosen item's details. To select an object on the plan, it can simply be clicked on.

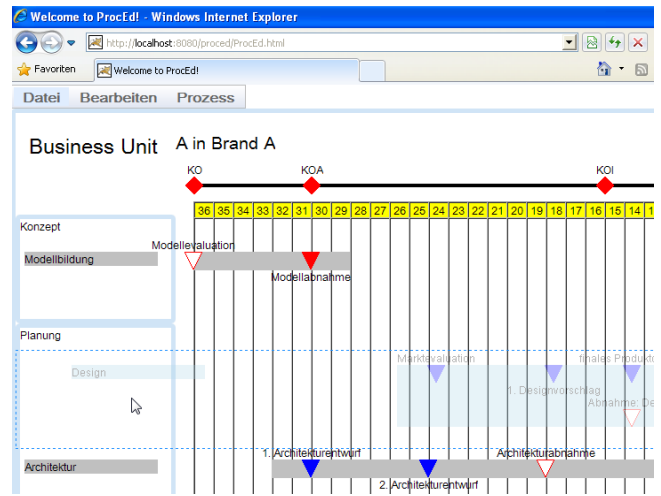

**Figure 2: Web-based dragging and dropping of items and collections of items.**

As input and output, an instance of the above defined grammar is used. After upload, the ASG is constructed from the file through the MontiCore-generated tools, although these objects are kept separate from the ones behind the displayed items. This separation enables us to replace both sides, grammar-generated objects and data-objects, without changing much at the corresponding one in case an engineering- or design-related update is necessary. No instance of an ASG-object is kept in a visualizable one and vice versa to achieve a very clean separation between the two worlds. Since the editor does not get any display-related information from the grammar, it has to decide itself on the positioning and use of visual elements such as icons, colors, etc.

To keep everything synchronized and to reduce computational effort and bandwidth, a central class containing a hashmap keeps all objects and links them to an icon-file that the user will finally see. This pattern makes searching and working in general with the data easier as if ASG-objects would keep their visualized counterparts themselves instead. A Command-pattern makes sure user-input is handled correctly and distributed to the right object and also adds an undo-/redo-functionality to the editor.

The web-application itself is secured through an SSL-connection and a required login provides a user-environment that lets a user keep a list of files he is working on. This part of the application could be extended, for example with functions like shared comments.

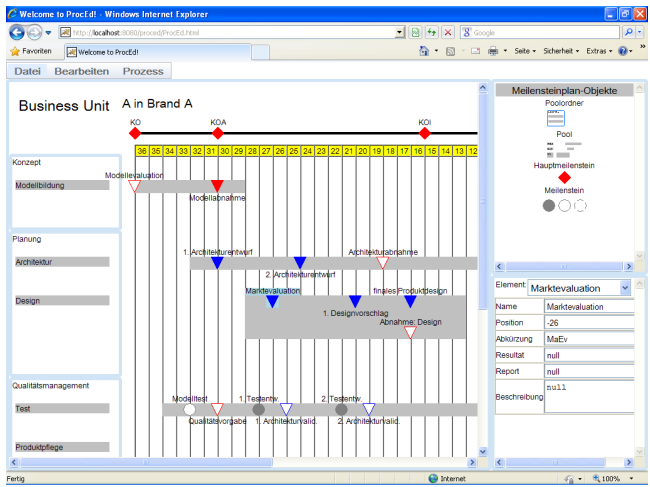

Figure 3: Screenshot of ProcEd, a web-based editor for instances of ProcDSLs.

## 5. APPLICATION AT AN AUTOMOTIVE MANUFACTURER

The already fixed layout of graphical representations of the corporate processes made questions about the application's appearance simple. The requirements gathered for this yielded to a login-screen, traditional file-menus, etc. Example process files in Microsoft Power Point like the one shown in Fig.4 specified the canvas' layout. The real difficulties therefore lay in the business-objects model and the different influences the classes have on each other.

We used a graphical representation of our model to be able to discuss it with selected domain-experts who had no education in computer science. UML was a good choice due to easily understandable class diagrams. However, we needed several iterations to get to a complete model-specification.

The clearly separated parts in the software though made it easy for us to keep up an agile workflow. While we could implement more and more of the interface, the model itself could be improved independently only requiring to generate a new lexer and parser using MontiCore to process the modified version, and correct a function call or the like in the separated part. Compared to a traditional approach which would have forced us to complete the model first and then build the application depending on it; using the aforementioned approach we used during development we simply could not only integrate but also embrace changes desired by the customer [1].

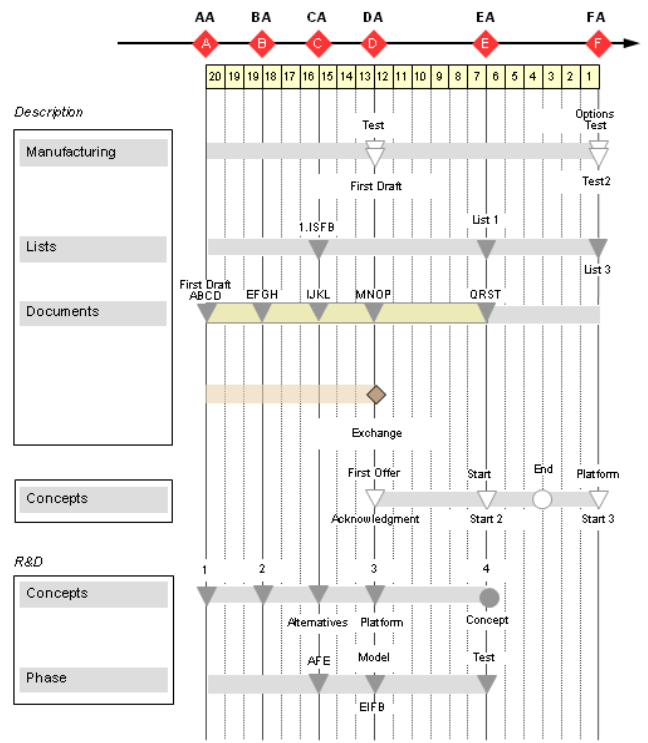

Figure 4: Elements for process description at a German automotive manufacturer.

## 6. RELATED WORK

The usage of DSLs in web-applications has received increased interest in the last years. But as outlined in [10] or [11], these activities did only focus on modeling an application's architecture and related workflows.

Our approach is different, because we did not use the DSL to define workflows, but to get a data exchange format that also serves as business layer in the resulting application. A change in the DSL would not result in a whole new application layout, only in differently working interfaces, leaving the former with the customer agreed on GUI intact. This is crucial, since the organizational layout for process documents is already company-wide communicated and approved. To understand the meaning behind different layers and associations though can be a tough job which needs flexible tools that will only change parts of the code that need to be changed with minimum effort.

The choice of Google's GWT as the framework used for realizing the web-editor was based on the excellent Eclipse integration and the number of features it includes. However, the most important factor, compared for example to [3], was the fact that GWT enables the developer to work in plain Java without having to care about data exchange or even JavaScript on the user's end. This is clearly an advantage because it decreases development time and simplifies testing and source code documentation. Moreover, as already discussed earlier, MontiCore generates the classes representing the grammar's ASG in Java which could be easily integrated with GWT.

Reasons a new DSL was used instead of implementing one of the business modeling languages available were the very company-specific process-layout on the one hand and the missing or incomplete formal specification of those languages on the other hand. As for example [13] notice, the Business Process Modeling Language (BPMN) lacks several concepts, like sub-processes with more than one instance, is partially ambiguous in its definition and has an incomplete mapping to the formal representation WS-Business Process Execution Language (BEPL). Moreover, BPMN did not let us represent the company-specifics we needed to be able to model, like a milestone's different meanings defined by the way different layers access it. This was a crucial fact since we needed to be able to display different views from different layers of the company onto the same sets of milestones and the connections between them. If one milestone changes, it has to be updated in every representation. This can only be achieved with a data-model representing exactly the company's structure.

## 7. CONCLUSION

In this paper, a formal, textual-based domain specific language for defining workflows was presented. Using this language, both documentation and modeling of organizational processes of a company is supported and also given instances of the DSL representing several workflows can be inspected.

For supporting a process engineer in modeling, documenting, and integrating different workflows, a graphical web-based editor using modern web 2.0 technologies was provided. Using this editor, workflows can be transparently presented and updated nearly everywhere in a company using a web-browser with state-of-the-art technologies already built-in.

The main contributions of this work – the formal description of organizational workflows on the one hand, and transparent access to the DSL's instances nearly everywhere on the other hand – provide valuable support for a process engineer's daily work. Furthermore, formal and machine-processable analysis of the DSL's instances can be realized both to check currently implemented workflows and to simulate changes in a company's processes to perform what-if-analysis.

With the MontiCore framework and toolkit, it was easy and efficient to define and implement the DSL-part of the editor, including a lexer, a parser, ASG classes, and standard context conditions. This and other examples from different domains have shown that the MontiCore infrastructure provides efficient techniques to develop DSL-based tools.

## 8. REFERENCES

[1] K. Beck, M. Beedle, A. van Bennekum, A. Cockburn, W. Cunningham, M. Fowler, J. Grenning, J. Highsmith, A. Hunt, R. Jeffries, J. Kern, B. Marick, R. C. Martin, S. Mellor, K. Schwaber, J. Sutherland, and D. Thomas. Manifesto for the Agile Software Development, 2001.
[2] BOC ADONIS http://www.boc-group.com/.
[3] Echo3 http://echo.nextapp.com/site/echo3.
[4] Google Web Toolkit http://code.google.com/intl/de/webtoolkit/.
[5] H. Grönniger, H. Krahn, B. Rumpe, M. Schindler, and S. Völkel. MontiCore: A Framework for the Development of Textual Domain Specific Languages. In *30th International Conference on Software Engineering (ICSE 2008), Leipzig, Germany, May 10-18, 2008, Companion Volume*, pages 925–926, 2008.
[6] IDS Scheer ARIS Business Designer http://www.ids-scheer.de/de/ARIS_ARIS_Platform/7796.html.
[7] H. Krahn, B. Rumpe, and S. Völkel. Efficient Editor Generation for Compositional DSLs in Eclipse. In *Proceedings of the 7th OOPSLA Workshop on Domain-Specific Modeling 2007*, 2007.
[8] H. Krahn, B. Rumpe, and S. Völkel. Integrated Definition of Abstract and Concrete Syntax for Textual Languages. In *Proceedings of Models 2007*, 2007.
[9] H. Krahn, B. Rumpe, and S. Völkel. Monticore: Modular development of textual domain specific languages. In *Proceedings of Tools Europe*, 2008.
[10] M. Nussbaumer, P. Freudenstein, and M. Gaedke. The impact of DSLs for assembling web applications. *Engineering Letters*, 13(3):387–396, 2006.
[11] M. Nussbaumer, P. Freudenstein, and M. Gaedke. Towards DSL-based web engineering. In *WWW '06: Proceedings of the 15th international conference on World Wide Web*, pages 893–894, New York, NY, USA, 2006. ACM.
[12] L. D. Paulson. Building Rich Web Applications with Ajax. *Computer*, 38(10):14–17, 2005.
[13] P. Wohed, W. M. van der Aalst, M. Dumas, A. H. ter Hofstede, and N. Russell. *Business Process Management*, chapter On the Suitability of BPMN for Business Process Modelling, pages 161–176. Springer Berlin, 2006.